\documentclass[12pt,superscriptaddress]{article}
\usepackage{palatino,cite,epsfig,amsmath,amssymb,color}

\newcommand{\ra}{\mbox{$\to$}}
\newcommand{\ee}{\mbox{${e}^+ {e}^-$}}
\newcommand{\bb}         {\mbox{${b}\bar{b}$}}
\newcommand{\ttbar}      {\mbox{${t}\bar{t}$}}
\newcommand{\glgl}       {\mbox{${gg}$}}
\newcommand{\mH}         {\mbox{$m_{H}$}}
\newcommand {\Ho}        {\mbox{${H}^{0}$}}

\newcommand{\WW}         {\mbox{${W}^+{W}^-$}}
\newcommand{\WWn}        {\mbox{${WW}$}}
\newcommand{\sqrts}      {\mbox{$\sqrt{s}$}}
\newcommand{\fb}         {\mbox{$\mathrm{fb}^{-1}$}}

\def\lsim{\mathrel{\raise.3ex\hbox{$<$\kern-.75em\lower1ex\hbox{$\sim$}}}}
\def\gsim{\mathrel{\raise.3ex\hbox{$>$\kern-.75em\lower1ex\hbox{$\sim$}}}}

\begin{document}

\begin{center}

{\Large\bf Model Independent Determination of the \\ 
Top Yukawa Coupling from LHC and LC}

\vspace{1cm}

{\sc 
K. Desch$^{a}$\footnote{Klaus.Desch@desy.de}, 
M. Schumacher$^{b}$\footnote{Markus.Schumacher@physik.uni-bonn.de}
}

\vspace*{1cm}

{\sl
$^a$ Institut f\"ur Experimentalphysik, Universit\"at Hamburg,
  Luruper Chaussee 149, \\ D-22761 Hamburg, Germany

\vspace*{0.3cm}

$^b$ Physikalisches Insitut der Universit\"at Bonn,
  Nussallee 12, \\ D-53112 Bonn, Germany

}

\end{center}

\begin{abstract}

We show how a measurement of the process $pp\to t\bar{t}H + X$ at
the LHC and a measurement of the Higgs boson branching ratios 
BR($H\to b\bar{b}$) and BR($H\to W^+W^-$) at a future linear
electron positron collider can be combined to extract
a model-independent measurement of the top quark Yukawa coupling.
We find that for 120 GeV $< m_H <$ 200 GeV a measurement precision of
15\% including systematic uncertainties
can be achieved for integrated luminosities of 300~\fb at the LHC 
and 500~\fb at the LC at centre-of-mass energy of 350 GeV.

\end{abstract}

\section{Motivation}

The Yukawa coupling of the Higgs boson to the heaviest quark, the top quark,
is of great interest for the study of the nature of electroweak symmetry
breaking and the generation of masses. While the Yukawa couplings to bottom
and charm quarks and to tau leptons and muons are in principle accessible
through the Higgs boson decay branching ratios, the Higgs boson decay into 
top quark pairs is kinematically forbidden for light Higgs bosons as they
are favoured by theory and electroweak precision data. The only 
Standard Model process that probes the top Yukawa coupling at tree 
level is the associated production of a \ttbar\ pair 
with a Higgs boson. This process occurs at the LHC (mainly 
\glgl\ra\ttbar\Ho ) as well as at the 
LC ( \ee\ra\ttbar\Ho ).
In the latter case the cross section is only significant at 
centre--of--mass energies in excess of 800~GeV. At the LHC, the final
states that have been investigated so far are \ttbar\bb\
\cite{lhcbb,Dai:1993gm,lhctop,cmstth} 
and \ttbar\WW \cite{lhcww,mrw}, the $\ttbar\tau^+\tau^-$ final state
is under study~\cite{eilam, ito}.
At tree level, their production rates are proportional to
%KD exchanged g by BR
$g_{ttH}^2\,\mathrm{BR}(\Ho\to\bb) $ and 
$g_{ttH}^2\,\mathrm{BR}(\Ho\to\WW) $, respectively. The
absolute values of $\mathrm{BR}(\Ho\to\bb)$ and $\mathrm{BR}(\Ho\to\WW)$ 
can be measured
accurately in a model independent way at the LC from the corresponding
decay branching ratios \cite{Aguilar-Saavedra:2001rg}. These can be
measured already at a first phase of the LC (\sqrts\ between 350 and
500 GeV). Thus, the combination of the measurements of both machines
can be used to determine the value of $g_{ttH}$ without model
assumptions and presumable before a second phase of the LC
($\sqrts\sim 1 $ TeV) would come into operation.

\section{Measurements at the LHC}

%\noindent
%\underline{LHC}\\
%
The results from the following ATLAS analyses of the \ttbar\Ho\ process
are used:

1. \ttbar\Ho\ with \ttbar\ra bbq$\ell\nu$ and \Ho\ra\bb\ ~\cite{lhcbb};

2. \ttbar\Ho\ with \Ho\ra\WW\ and two like-sign leptons~\cite{lhcww};

3. \ttbar\Ho\ with \Ho\ra\WW\ and three leptons~\cite{lhcww}.

\begin{table}[h!]
\begin{center}
\scalebox{0.9}{
\begin{tabular}{|c||c|c||c|c|} \hline
\mH & \multicolumn{2}{c||}{30\fb} & \multicolumn{2}{|c|}{300\fb} \\ \cline{2-5}
(GeV) & \ttbar\Ho\, \Ho\ra\bb\ & background & \ttbar\Ho\, \Ho\ra\bb\ &
      background \\ \hline
100   & 83.4 &  303.4 & 279.0 & 1101.3  \\ \hline
110   & 63.0 &  275.7 & 232.5 & 1140.6  \\ \hline
120   & 43.0 &  234.1 & 173.1 & 1054.2  \\ \hline
130   & 26.5 &  200.1 & 112.5 & 1015.8  \\ \hline
140   & 13.9 &  178.2 &  62.4 & 947.1  \\ \hline
\end{tabular}
}
\end{center}
\caption{ \label{tab:lhcbb} Expected number of signal and background events
for the \ttbar\Ho\ with \ttbar\ra bbq$\ell\nu$ and \Ho\ra\bb\ analysis
at LHC ~\cite{lhcbb}.}
\end{table}
\begin{table}[h!]
\begin{center}
\scalebox{0.9}{
\begin{tabular}{|c||c|c||c|c||c|c||c|c||} \hline
 & \multicolumn{4}{|c||}{30 \fb} & \multicolumn{4}{|c|}{300 \fb} \\ \cline{2-9}
\mH  & \multicolumn{2}{|c||}{\ttbar\Ho\,\Ho\ra\WWn (2$\ell$)}
      & \multicolumn{2}{|c||}{\ttbar\Ho\,\Ho\ra\WWn (3$\ell$)}
      & \multicolumn{2}{|c||}{\ttbar\Ho\,\Ho\ra\WWn (2$\ell$)}
      & \multicolumn{2}{|c||}{\ttbar\Ho\,\Ho\ra\WWn (3$\ell$)} \\ \cline{2-9}
(GeV)  & signal& bckgr  
  & signal& bckgr  
  & signal & bckgr  
  & signal & bckgr  \\ \hline
120 & 4.4  & 19.6 &  2.7 & 21.2 & 12.7 & 80.6 &  10.5 & 97.6 \\ \hline
140 & 15.0 & 19.6 &  8.7 & 21.2 & 50.0 & 80.6 &  33.7 & 97.6 \\ \hline
160 & 21.1 & 19.6 & 13.0 & 21.2 & 72.3 & 80.6 &  55.3 & 97.6 \\ \hline
180 & 17.3 & 19.6 & 10.3 & 21.2 & 60.9 & 80.6 &  41.7 & 97.6 \\ \hline
200 & 10.5 & 19.6 & 5.7  & 21.2 & 43.2 & 80.6 &  26.4 & 97.6 \\ \hline
\end{tabular}
}
\end{center}
\caption{ \label{tab:lhcww} Expected number of signal and background events
for the \ttbar\Ho\ with \Ho\ra\WW\ (two like-sign leptons and three
leptons, respectively) analyses at LHC~\cite{lhcww}.}
\end{table}

The expected numbers of selected signal and background events in the
three channels for various Higgs masses and total integrated luminosities 
of 30~\fb\, and 300~\fb\, are
listed in Tables~\ref{tab:lhcbb} and \ref{tab:lhcww}.  The results
obtained in this sub-section are based on the anticipated data sample of
\textit{one} LHC detector, with the luminosity per detector quoted above.

From the expected event numbers we first estimate the 
uncertainty (statistical and systematic) on the measured cross
section $\sigma_{tth}^{data}$. Further uncertainties arise when
$\sigma_{tth}^{data}$ is compared to the theoretical prediction
as a function of $g_{tth}$.

The uncertainty on the observed cross section $\sigma_{tth}^{data}$
is calculated as
\begin{eqnarray*}
(\Delta\sigma_{tth}^{data}/\sigma_{tth}^{data})^2 & = &
(S+B) / S^2 + 
 (\Delta B_{syst})^2 / S^2 + 
 (\Delta {\cal L})^2 / {\cal L}^2  +
 (\Delta {\epsilon})^2 / {\epsilon}^2.
\end{eqnarray*}

Here, $S (B)$ is number of signal (background) events. $\Delta
B_{syst}$ is the uncertainty on the background determination from
sideband data (10\% in the $h\to\bb$ channel at high luminosity, 5\% otherwise).
$\Delta {\cal L}$ is the error on the integrated luminosity
(5\%) and $\Delta \epsilon$ is the error on the determination of the 
efficiency. This error involves uncertainties on the tagging
efficiency for individual b-jets (3\%) and leptons (3\% from isolation
requirement and 2\% from reconstruction efficiency) and an overall
detector efficiency uncertainty of 2\% (following ~\cite{due}). The total value
of $\Delta \epsilon$ is then calculated for each channel individually depending 
on the number of leptons and b-jets.

The expeced error including systematic uncertainties and 
taking into account only the statistical error of each channel is 
shown in Table~\ref{tab:lhc2}. For the \Ho\ra\WW decay mode the signal and background
from the two lepton and three
lepton channels are added together since their signal contributions are 
exclusive and the overlap in the background is small.
The obtained result is consistent with the study presented in~\cite{due}.

\begin{table}[t!]
\begin{center}
\begin{tabular}{|c||c|c||c|c|} \hline
\mH\ & \multicolumn{2}{|c||}{30 \fb} & \multicolumn{2}{|c|}{300
  \fb} \\ \cline{2-5}

(GeV) & \Ho\ra\bb & \Ho\ra\WWn\ &\Ho\ra\bb & \Ho\ra\WWn\   \\ \hline
100   & 0.398(0.236)    &              & 0.249(0.133)&  \\ \hline
110   & 0.476(0.292)    &              & 0.287(0.159)&  \\ \hline
120   & 0.598(0.387)    & 1.023(0.974) & 0.345(0.202)& 0.732(0.611) \\ \hline
130   & 0.840(0.568)    & 0.524(0.492) & 0.488(0.299)& 0.362(0.295) \\ \hline
140   & 1.444(0.997)    & 0.370(0.339) & 0.804(0.509)& 0.252(0.193) \\ \hline
160   &                 & 0.287(0.254) &             & 0.196(0.137) \\ \hline
180   &                 & 0.331(0.300) &             & 0.221(0.163) \\ \hline
200   &                 & 0.486(0.454) &             & 0.282(0.222) \\ \hline

\end{tabular}
\end{center}
\caption{ \label{tab:lhc2} Expected relative precision
on $\sigma_{ttH} \times BR(H\to X)$ for the various LHC \ttbar\Ho\
analyses including systemtatic uncertainties (statistical error only). 
For $\Ho\ra\WW$~ the expected signal and background in the two and
three lepton final state have been added.}
\end{table}

In the next step the uncertainty on $g_{ttH}^2 * BR(H\to\bb/WW)$ which
arises when the observed $\sigma_{ttH} * BR(H\to \bb/WW)$ is compared to its
theoretical prediction. These uncertainties consist of the
uncertainties in the proton structure functions (5\%~\cite{sf,pdf}) and
uncertainties in the calculation of the production cross section.
Recent full
NLO calculations estimate the uncertainty of the cross section prediction
to be approximately 15\% from a variation of the hard 
scale~\cite{ref1_nlo_tev,ref1_nlo_lhc,ref2_nlo_tev,ref2_nlo_lhc}.
The total theoretical uncertainty $\Delta\sigma_{ttH}^{theo}$ is obtained by
adding the above two sources in quadrature.

Finally, the total uncertainty  $\Delta(g_{ttH}^2 * BR(H\to\bb/\WW))$ is
obtained according to
\begin{eqnarray*}
& \Delta(g_{ttH}^2 * BR(H\to\bb/\WW))^2/(g_{ttH}^2 * BR(H\to\bb/\WW)^2
& =  \\
& (\Delta\sigma_{ttH}^{theo})^2/(\sigma_{ttH}^{theo})^2 + 
(\Delta\sigma_{ttH}^{data})^2/(\sigma_{ttH}^{data})^2. & 
\end{eqnarray*}

%\vspace{2em}
\section{Measurements at the LC}

%\noindent
%\underline{LC}\\

At the LC, the decay branching ratios into b quark pairs and W boson pairs
can be measured at \sqrts = 350~GeV to the precision listed in 
Table~\ref{tab:lc}\cite{Aguilar-Saavedra:2001rg}. The precise 
model-independent measurement of Higgs boson branching ratios exploits
the measurement of the Higgs-strahlung process $e^+e^-\ra HZ$. 
Events from specific Higgs decays, e.g.~$H\to\bb$ and $H\to\WW$ can be
cleanly identified. The branching ratio is determined by normalizing the
observed rate for a specific Higgs decay to
the total Higgs-strahlung rate. The latter
can be measured from the selection of $Z\ra\ell\ell$
events where the invariant mass of the recoil system is consistent with the
Higgs mass, independent of the Higgs decay.

\begin{table}[b!]
\begin{center}
\begin{tabular}{|c|c|c|} \hline
\mH~(GeV)     & $\Delta$BR(\bb)/BR(\bb)   & $\Delta$BR(WW)/BR(WW) \\ \hline
100     & 0.024                   &                      \\ \hline
120     & 0.024                   & 0.051                 \\ \hline
140     & 0.026                   & 0.025                 \\ \hline
160     & 0.065                   & 0.021                 \\ \hline
200     &                         & 0.021                 \\ \hline
\end{tabular}
\end{center}

\caption{\label{tab:lc} 
Relative precision on the branching ratio for $\Ho\to\bb$
and $\Ho\to\WW$ expected
for a LC running at \sqrts = 350~GeV with 500 fb$^{-1}$.}

\end{table}

\section{Results}

\begin{figure}[htbp]
\centerline{
\epsfig{file=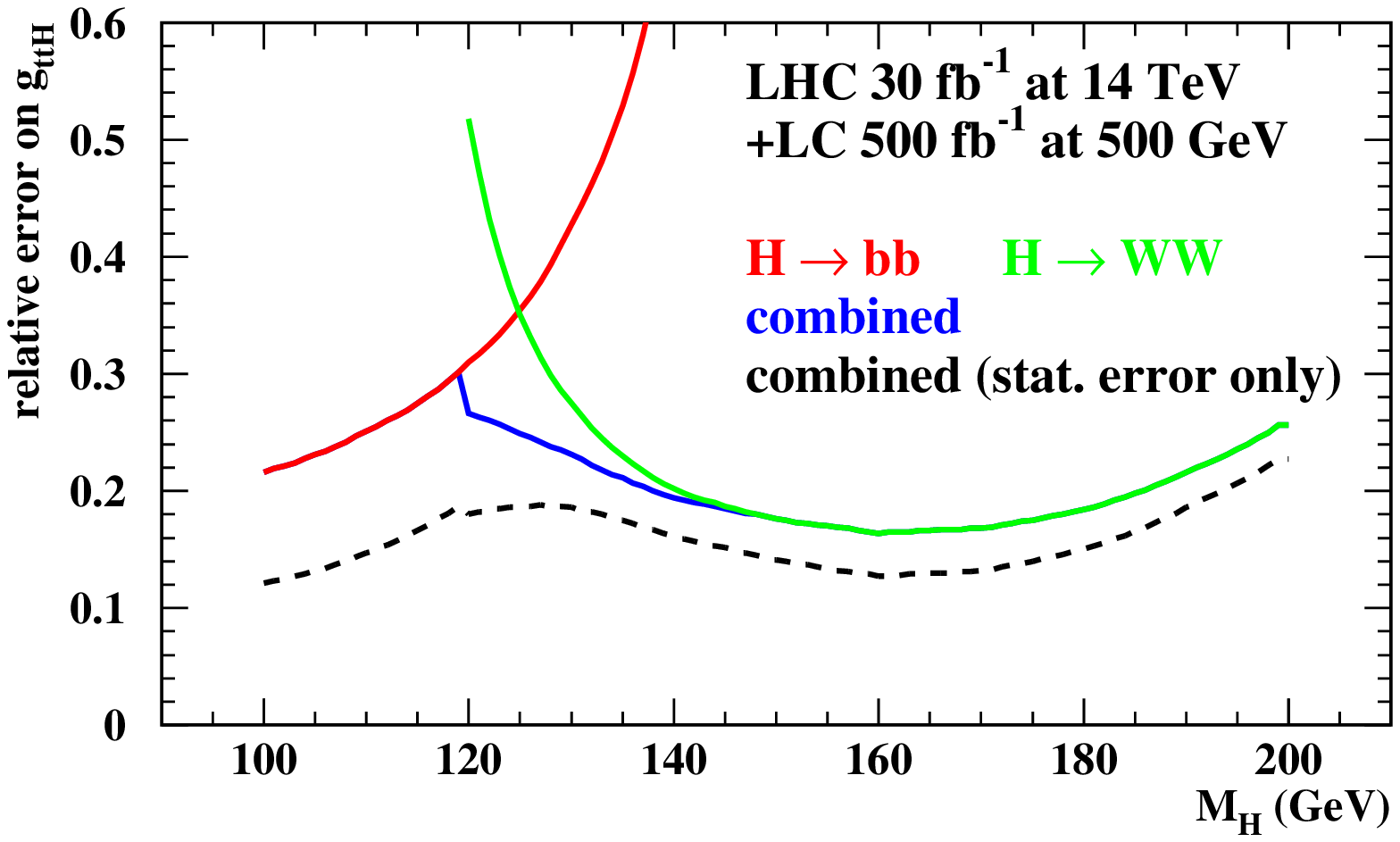,width=0.485\linewidth}
\epsfig{file=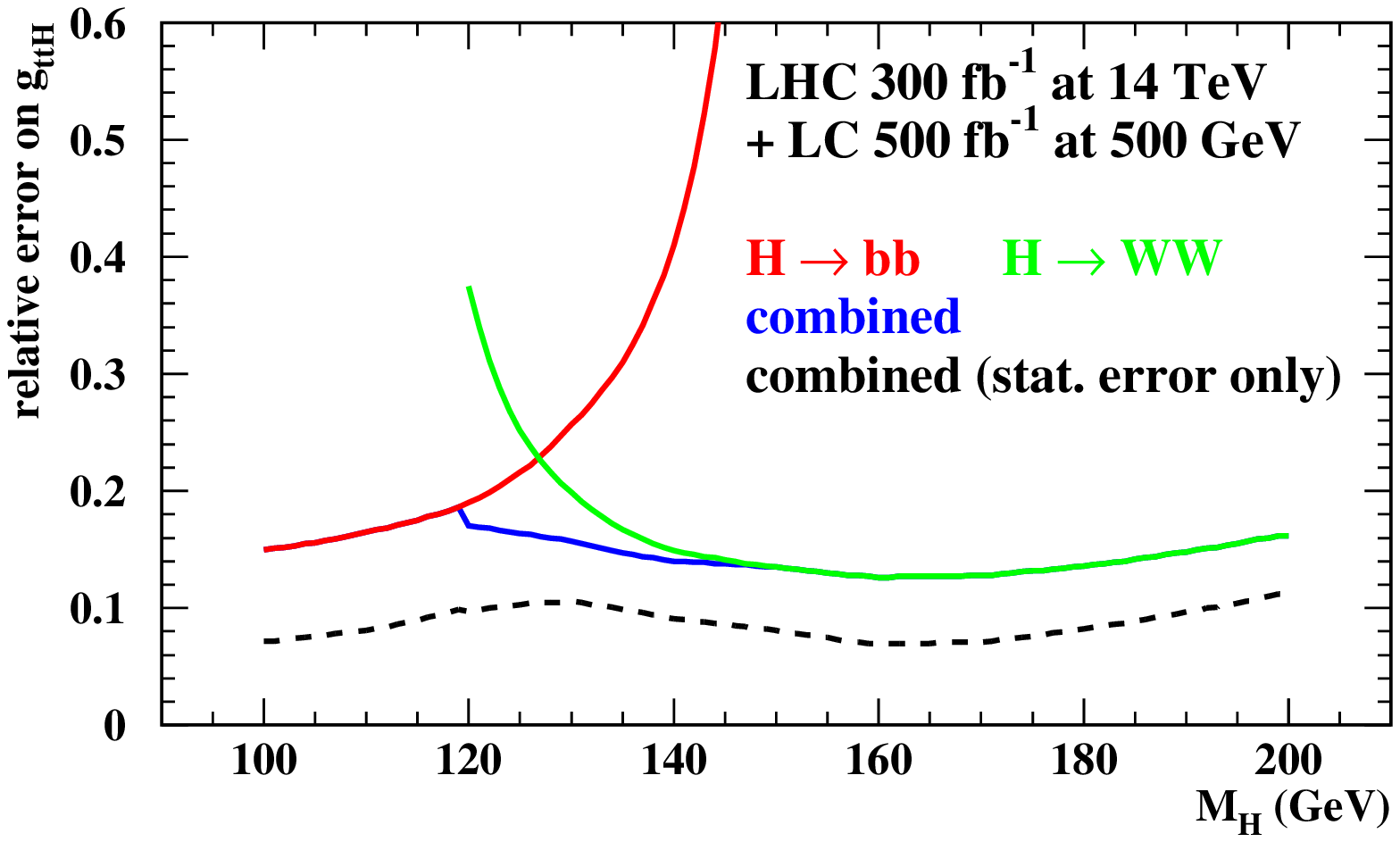,width=0.485\linewidth}  }
\caption{\label{fig:result1} 
Achievable precision on the top Yukawa
coupling from 30~\fb at the LHC and 500~\fb at the LC (left),
and from 300~\fb at the LHC and 500~\fb at the LC
(right). The red curve shows precision obtainable from the 
\Ho\ra\bb\ final state, the green from the \Ho\ra\WW\ final state and
the blue curve from the combination of the two. The dashed lines 
show the expected precision taking into account only statistical errors.}
\vspace{0.2in}
\end{figure}

For an extraction of the top quark Yukawa coupling at each Higgs mass
we combine the LHC rate measurement of \ttbar\Ho with \Ho\ra\bb\ or
\Ho\ra\WW\ with the corresponding measurement of the branching ratio
at the LC.  We make the tree level assumption that the cross section
$\sigma_{ttH}$ is proportional to $g_{ttH}^2$. Thus, the relative
error on $g_{ttH}$ is simply given by $\Delta g_{tth} / g_{ttH} = 0.5
\Delta \sigma_{ttH} / \sigma_{ttH}$. The relative error on
$\sigma_{ttH}$ is obtained by adding in quadrature the statistical
and systematic uncertainties as
described above and the error of the LC branching ratio
measurement. The combination of the \bb\ and \WW\ final states is
performed by
\begin{equation*}
\left( \frac{\Delta g_{tth}} {g_{ttH}} \right)_{comb.}^{-2}= 
\left( \frac{\Delta g_{tth}} {g_{ttH}} \right)_{WW}^{-2} +
\left( \frac{\Delta g_{tth}} {g_{ttH}} \right)_{b\bar{b}}^{-2}. 
\end{equation*}
The relative accuracies on the top quark Yukawa coupling achievable are
summarised in Table~\ref{tab:result}. In Figure~\ref{fig:result1} the 
relative accuracy from the $H\to\bb$ and $H\to WW$ channels are
shown individually and combined both for low and high luminosity at 
the LHC. Also shown is the results which would be obtained when all
systematic errors were neglected. 

%%%%%%%%%%%%%%%%%%%%%%%%%%%%%%%%%%%%%%%%%%%%%%%%%%%%%%%%%%%%%%
\begin{table}[htbp]
\begin{center}
\begin{tabular}{|c||c|c|c||c|c|c|} \hline
\mH  & \multicolumn{3}{|c||}{30 \fb} & \multicolumn{3}{|c|}{300
  \fb} \\ \hline
(GeV) & bb & WW & bb+WW& bb & WW & bb + WW \\ \hline
100  &  0.22(0.12) &  & & 0.15(0.07) &  & \\ \hline                              
110  &  0.25(0.15) &  & & 0.17(0.08) &  & \\ \hline                              
120  &  0.31(0.19) &  0.52(0.49)  & 0.27(0.18) & 0.19(0.10) &0.38(0.31) & 0.17(0.10) \\ \hline
130  &  0.43(0.28) &  0.28(0.25)  & 0.23(0.19) & 0.26(0.15) &0.20(0.15) & 0.16(0.11) \\ \hline
140  &  0.72(0.50) &  0.20(0.17)  & 0.19(0.16))& 0.41(0.26) &0.15(0.10) & 0.14(0.09) \\ \hline 
150  &              & 0.18(0.14)  &              & 1.88(1.21) &0.14(0.08) & 0.14(0.08) \\ \hline
160  &              & 0.16(0.13)  &              &              &0.13(0.07) &  \\ \hline
170  &              & 0.17(0.13)  &              &              &0.13(0.07) &  \\ \hline
180  &              & 0.18(0.15)  &              &              &0.14(0.08) &  \\ \hline
190  &              & 0.22(0.19)  &              &              &0.15(0.10) &  \\ \hline
200  &              & 0.26(0.23)  &              &              &0.16(0.11) &  \\ \hline
\end{tabular}
\end{center}
\caption{\label{tab:result} Expected relative error on the top 
Yukawa coupling $g_{ttH}$ from the rate measurement including all
systematic uncertainties (statistic errors only) at the LHC  
and from the branching ratio measurement at the LC.}
\end{table}

\begin{figure}[hbtp]
\centerline{
\epsfig{file=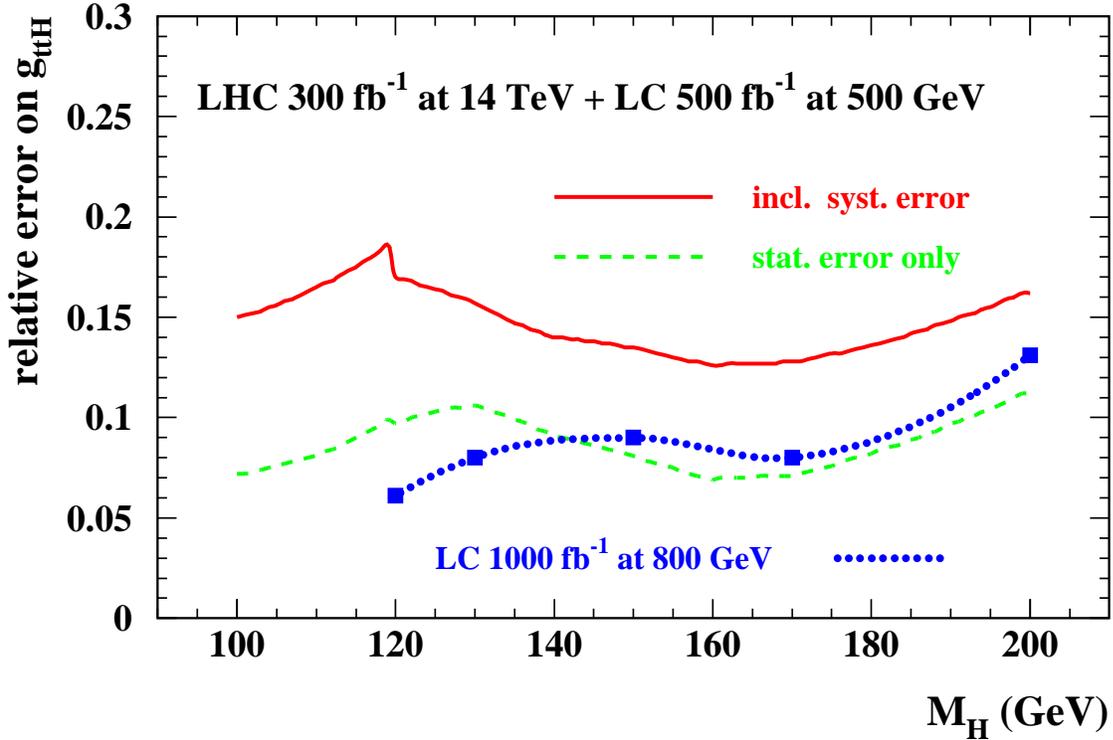,width=0.92\linewidth}  }
%KD changed
\caption{\label{fig:result2} Achievable precision on the top Yukawa
coupling from 300~\fb at the LHC and 500~\fb at the LC at 350 GeV
taking into account all systematic uncertainties (solid red curve)  
and  using only statistical errors (dashed green curve) .
For comparison the expected precision from 1000~\fb at the LC at 800
GeV alone (dotted blue curve) is also shown.}
\end{figure}

For 300~\fb at the LHC and 500~\fb at the LC
the obtainable relative uncertainty is approximately 15\% for a Higgs
boson mass between 120 and 200 GeV. The purely statisical uncertainty
ranges from 7\% to 11\% as shown in Fig.~\ref{fig:result2}.
The size of the obtained uncertainties is comparable to those obtained
for the LHC alone~\cite{due} but in contrast to the latter no
model-dependent assumptions are made.
%KD added
In Fig.~\ref{fig:result2} we also show the precision which can be achieved
at the LC alone if operated at 800~GeV center-of-mass energy~\cite{gay}
from the measurement of the \ee\ra\ttbar\Ho\ process with \Ho\ra\bb\ and
and \Ho\ra\WW\ combined.

\section*{Acknowledgements}

We would like to thank Jochen Cammin and Arnaud Gay for valuable discussions.

\end{document}